# A Study on the Improvement of Code Generation Quality Using Large Language Models Leveraging Product Documentation

Takuroh Morimoto[1] and Harumi Haraguchi[2*]

[1]Department of Information Engineering, Faculty of Engineering, Ibaraki University, Hitachi, JAPAN
[2]Graduation School of Science and Engineering, Ibaraki University, Hitachi, JAPAN
*Corresponding harumi.haraguchi.ie@vc.ibaraki.ac.jp

**ABSTRACT**
Research on utilizing Large Language Models (LLM) in system development is actively expanding, particularly in automated code generation and test generation. Among various testing approaches, End-to-End (E2E) testing is essential for ensuring application quality. However, while research on test code generation has predominantly focused on unit tests, there is limited research on E2E test code generation. This study proposes a method to automatically generate E2E test code using LLM based on product documentation. Product documentation includes manuals, tutorials, FAQs, and step-by-step guides that help users accomplish their tasks within an application. While the importance of E2E testing in software development is increasing, the significant effort required to create and maintain test code remains a challenge. This research aims to improve the coverage and quality of E2E test code by leveraging the detailed instructions in the product documentation. The proposed method takes product documentation as input and generates E2E test cases and corresponding test codes using tailored prompts for LLM. This two-step approach enables the model to accurately interpret the intent of the documentation and transform it into executable test code. Experiments were conducted using a web application with six major functionalities: authentication, profile management, and discussion features. The generated test code was evaluated by comparing product documentation, requirement specifications, and user stories. Evaluation metrics included the percentage of successfully compiled tests and functional coverage. The results demonstrated that test code generated from product documentation achieved high functional coverage, particularly in authentication, discussion, commenting, and user management features. Compared to other document types, it consistently produced high-quality test code. These findings suggest that leveraging product documentation can lead to higher-quality E2E test code, ultimately improving overall software quality.

## 1. Introduction
### 1.1. Background

In software development, it has been reported that code quality significantly impacts the development period and the number of bugs [1]. In particular, continuing to add features and release software while maintaining low code readability and maintainability has been pointed out to not only drastically increase the effort required for modifications but also increase the uncertainty of quality. Improving code quality is essential to making the most of limited human resources. In fact, research findings suggest that maintaining high code quality significantly reduces defect rates and greatly improves development speed [1]. On the other hand, there are various aspects of software code quality, among which testability is considered particularly important [2] [3]. Highly testable code allows for smooth defect detection and functional verification, ultimately reducing maintenance costs and minimizing rework before release. To enhance these quality attributes, simple refactoring and adherence to coding standards alone are insufficient; a comprehensive system-level testing process is crucial.

Representative types of software testing include unit testing, integration testing, and End-to-End (E2E) testing, with E2E testing being particularly important as it can evaluate user behavior [4]. However, modern web applications and cloud services involve complex interactions between multiple layers, such as frequent UI modifications and external API integrations, leading to a strong tendency for test cases to increase with each release. Consequently, the workload required for testing cannot be underestimated, and challenges such as the burden of rewriting tests to accommodate specification changes and the inefficiency of repeatedly conducting manual regression testing remain significant issues [5].

Recently, the application of large language models (LLM) to the software testing phase has been advancing. As a test automation technology, research primarily focuses on automatically generating unit tests from code or requirements and improving unit tests using LLM [6] [7] [8]. Regarding E2E testing, there have been studies on generating test scenarios based on the characteristics of web applications, but research on applying LLM to the automatic generation of E2E test code is still limited [9]. Meanwhile, the practical implementation of LLM in the E2E testing field is progressing, and software that can automatically execute E2E tests based on simple natural language input or generate E2E test code from existing code has been introduced [10] [11]. By adopting such software, it is expected that the repetitive, labor-intensive nature of testing can be reduced, allowing development teams to focus more on new features and maintenance tasks.

Regarding LLM-based automatic code generation, some methods generate code from requirements, while others allow users to upload product specifications and interactively generate test cases and test scenarios. Such documentation plays a crucial role in improving the quality of LLM-generated code [12].

### 1.2. Objective

This study proposes a method for automatically generating E2E test code based on product documentation. By inputting user-facing product documentation to generate E2E test code, the proposed approach aims to reduce the effort required for developers and QA engineers to write tests from scratch while also improving maintainability when UI changes occur. Furthermore, this study demonstrates how documentation contributes to LLM-based code generation, ultimately promoting efficient software development using LLM.

### 1.3. Structure of This Paper

This paper is structured as follows. Chapter 2 provides an overview of related technologies. Chapter 3 discusses existing test code generation methods, including the flow of test code generation and the input materials utilized in code generation. Chapter 4 presents the proposed method, detailing the approach for generating E2E test code and the prompts used. Chapter 5 describes the experimental methodology and results. Finally, Chapter 6 provides a summary and conclusion.

## 2. Related Technologies
### 2.1 E2E Testing

From a user's perspective, End-to-End (E2E) testing verifies the entire system, including multiple components, external services, and databases. For example, a web application involves testing the entire flow from the login page to performing interactions with buttons and forms to use the service.

Since E2E testing is often conducted in production-like environments, it goes beyond simple, functional checks. It comprehensively ensures quality by covering UI interactions, integration with external APIs, and system behavior during network communication failures. On the other hand, E2E testing tends to have relatively high execution and maintenance costs. Unlike unit testing, modifying test cases often requires broad updates, making maintenance more complex. However, properly implementing E2E testing allows early detection of critical user workflows and functional integration issues, enabling developers to address problems proactively without waiting for end-user feedback.

### 2.2 Unit Testing

Unit testing is a testing method that verifies the smallest components of software, such as functions, methods, or classes. It primarily ensures that individual unit's function correctly as per specifications. Implementing unit testing improves early bug detection, enhances safety during refactoring, and contributes to overall development quality. Additionally, unit tests are relatively easy to automate, making them well-suited for integration into Continuous Integration (CI) environments. However, maintaining test code and ensuring adequate coverage comes with a cost. Therefore, designing software with testability in mind and prioritizing test coverage based on importance are critical aspects of an effective testing strategy.

### 2.3 Playwright

Playwright is an E2E testing framework developed by Microsoft that offers cross-browser support (Chromium, Firefox, WebKit) and features such as parallel test execution. Competitors include well-known frameworks such as Cypress and Selenium. However, Playwright stands out for its adaptability to modern technology stacks and ability to perform fast, headless browser testing. As shown in Figure 1, Playwright recorded a high number of downloads as of January 2025, showing a significant growth rate compared to traditional E2E testing frameworks such as Cypress and Nightwatch. Notably, its rapid adoption since late 2024 suggests increasing support from the developer community.

This study adopts Playwright as the E2E testing framework due to its strong recognition as an open-source E2E testing tool.

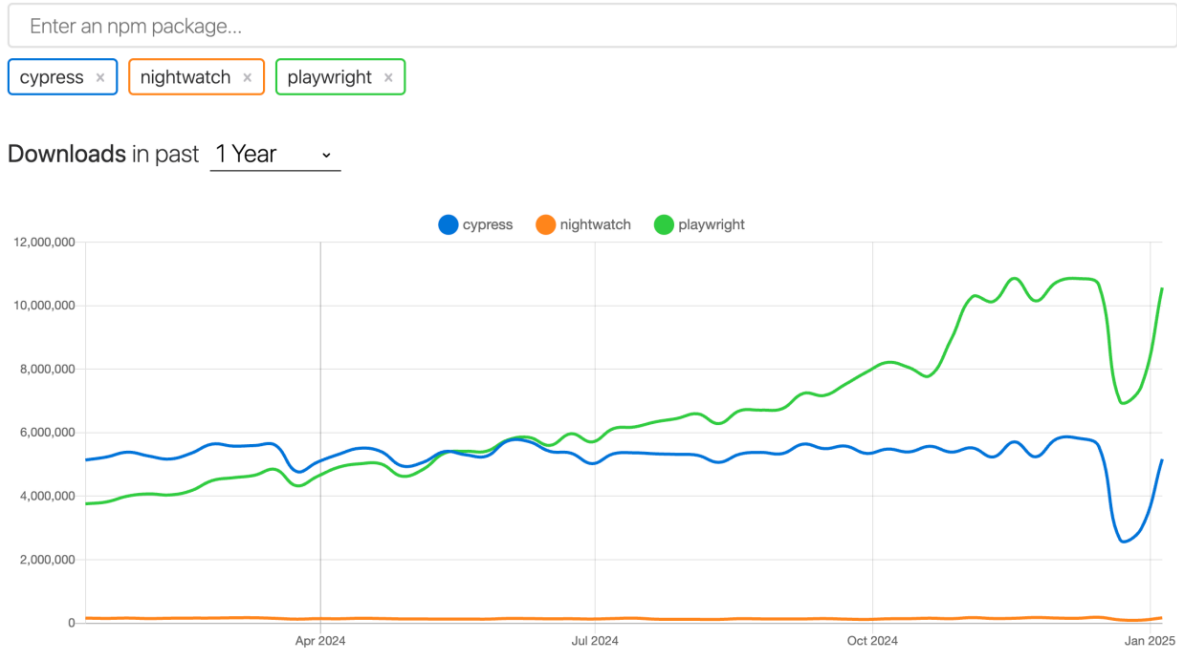

Figure. 1 Trends in npm downloads for major E2E testing frameworks (as of January 2025)

### 2.4 Product Documentation

In this study, product documentation refers to a collection of documents that organize the necessary features and procedures for using a tool or application, guiding users in achieving their goals. In the modern era, where products are becoming increasingly complex, the clarity of documentation is a crucial factor affecting user experience and product evaluation. Particularly in technology-related products, frequent additions and updates of new features necessitate continuous maintenance of up-to-date specifications and operation procedures. The product documentation includes instructions, common issues users may encounter, and best practices for using the software. By incorporating examples tailored to the software's usage context, screenshots, and short tutorial videos, users can quickly grasp essential information.

### 2.5 Requirement Specification Document

A requirement specification document outlines the business requirements, functional requirements, and non-functional requirements that a system must meet [13]. It defines the purpose, scope, and constraints of the system being developed, serving as an essential artifact for establishing consensus among stakeholders. The document details the required system functionalities, including screen requirements, report requirements, data requirements, and external interface requirements. Additionally, non-functional requirements such as performance and security considerations are also included, making the requirement specification document a fundamental reference for system development.

### 2.6 User Stories

A user story is an Agile development method for defining software requirements from an end-user's perspective. By avoiding technical jargon and using everyday language, user stories help development teams clearly understand the value they deliver to customers. Unlike detailed technical specifications, user stories focus on the goal to be achieved rather than implementation details. This approach allows development teams to fully grasp the value and problem to be solved before diving into the specifics of implementation.

## 3. Existing Test Code Generation Methods
### 3.1 Test Code Generation Workflow

Research on test generation utilizing LLM includes methods that generate unit tests as intermediate outputs from requirements before ultimately generating test code and methods that automatically generate test cases based on application characteristics using LLM [9] [12]. Other approaches involve extracting and generating test cases from requirement specification documents and design documents and then generating test code from those test cases [14]. A common feature among these studies and software tools is that they use requirements and design documents as input, generate test cases using LLM, and then generate test code from those test cases. However, these approaches do not utilize product documentation, leaving unresolved issues regarding generating test cases that accurately reflect actual user operation flows.

### 3.2 Input Data Used in Code Generation

In research and software related to code generation, requirement specification documents and user stories are commonly used. Requirement specification documents describe a system's overall functionality and specifications. In contrast, user stories are written from the user's perspective in a simplified format and are often used as input for LLM-based research on the automatic generation of specifications and requirements [15]. However, little research has been conducted on utilizing product documentation, which provides detailed descriptions of actual user operation procedures and concrete usage scenarios.

## 4. Proposed Method
### 4.1 E2E Test Code Generation

This study focuses on product documentation, which has not been utilized in previous test code generation research. It aims to develop an LLM-based E2E test code generation method using product documentation. The product documentation provides detailed descriptions of actual user operations and concrete usage scenarios, making it well-suited for generating test cases that accurately reflect fundamental user interactions required for E2E testing. Figure 2 shows the flowchart for generating E2E test cases from documentation using LLM, based on the conventional test code auto-generation workflow introduced in Section 3.1. As illustrated in Figure 2, the test code generation workflow in this study consists of the following steps:

Input Documents: Product documentation, requirement specifications, and other related documents are provided as input to the LLM, which then generates test cases. At this stage, the LLM identifies testable features based on operation procedures and use cases described in the documentation. It then constructs detailed test scenarios for each of these features.

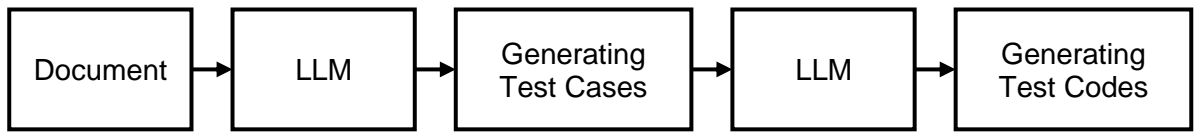

Figure. 2 The flow for generating E2E test code

Generating Test Code: The generated test cases are fed back into the LLM to produce the final executable test code. During this second phase, the test case contents are transformed into a format compatible with the Playwright framework. This two-stage LLM-based process successfully converts product documentation into executable test code.

By utilizing LLM with different roles at each stage, more accurate and practical test code can be generated. This approach aims to reduce the manual effort required to write test code while ensuring that the generated test cases are aligned with product specifications and of high quality. Additionally, since the test code can be automatically regenerated in response to updates in product documentation, this method is expected to improve test maintainability. Table 1 summarizes the roles of the two LLM used in this study.

Table 1 The role of LLM in the project

| LLM | Description |
| --- | --- |
| Test Case | Extract test cases from the document and generate them in JSON format. |
| Code Generation | Generate E2E test code using Playwright based on the extracted test cases. |

The JSON format output generated in the test case generation phase is shown in Figure 3.

```
1  {
2    "testCases": [
3      {
4        "title": "User purchases a product",
5        "description": "Verify standard purchase flow",
6        "steps": [
7          {
8            "action": "Access product listing page",
9            "expectedResult": "Product list is displayed"
10         }
11       ]
12     }
13   ]
14 }
```

Figure. 3 Source Code

The implemented web application used for evaluation is illustrated in Figure 4

```
Here's the Playwright e2e test code based on the provided test cases:

```typescript
import { test, expect } from '@playwright/test';

test.describe('Discussion Application E2E Tests', () => {
  test('User registration with new team creation', async ({ page }) => {
    await page.goto('/');
    await page.click('text=Register');

    await expect(page).toHaveTitle(/Registration/);

    await page.fill('input[name="firstName"]', 'John');
    await page.fill('input[name="lastName"]', 'Doe');
    await page.fill('input[name="email"]', 'john.doe@example.com');
    await page.fill('input[name="password"]', 'securePassword123');
    await page.fill('textarea[name="bio"]', 'I am a new user');

    await page.uncheck('input[name="chooseTeam"]');

    await page.click('button:has-text("Submit")');

    await expect(page).toHaveURL(/.*dashboard/);
  });
```

Figure. 4 Example of result

### 4.2 Prompt Design

As shown in Figure 2, this study employs LLM at two key points:
- Generating test cases from documentation
- Generating test code from test cases

The prompts used for instructing the LLM were designed based on official prompts published in existing software documentation that supports test code generation features [14]. Table 2 lists the prompts used in this study.

Minimizing the number of test cases is desirable for E2E tests since they require fewer test cases than unit and integration



tests [16]. Furthermore, as this study focuses on improving E2E test code quality, only standard test cases are generated, and error cases are not considered. To ensure fair comparisons in the experiments, only text-based product documentation is used in this study. However, product documentation often includes screenshots or other visual elements illustrating operation steps.

Therefore, this study utilizes Claude 3.5 Sonnet, a high-performance LLM developed by Anthropic, which is capable of reading images and extracting information accurately (API Model Name: claude-3-5-sonnet-20240620 [17]).This model is particularly well-suited for this research because of its high accuracy in extracting procedural and specification details from complex documents and its large context window, which allows for processing long documents in a single pass.

Table 2 Prompt

| Prompt Name | Description |
| --- | --- |
| Test Case | A dedicated prompt designed to analyze documents and generate structured E2E test cases in JSON format. This prompt utilizes natural language processing to identify and extract key user operation scenarios, transforming them into well-defined test specifications that include clear actions and expected outcomes. |
| Code Generation | A dedicated prompt designed to convert the analyzed test cases into a format executable by the Playwright framework. This prompt transforms the contents of the test cases into Playwright code, implements appropriate assertions, and generates complete test code. |

## 5. Evaluation Experiment
### 5.1 Experimental Method
#### 5.1.1 Benchmark

To evaluate E2E testing, a benchmark application was developed that includes functionalities commonly implemented in web applications. Table 3 lists the functions of this application. This application was developed by referring to the source code of open-source software [18]. Additionally, for the experiment, product documentation was created to illustrate the application's screen transitions and operation procedures.

Table 3 List of Application Features

| Prompt Name | Description | Function List |
| --- | --- | --- |
| Authentication | Login and registration | • User registration (name, email, password, team name)<br>• Login (email, password)<br>• Logout |
| Profile | User profile management | • View profile<br>• Update profile |
| Discussion | Team discussion management | • Create discussion<br>• View discussion<br>• Update discussion<br>• Delete discussion |
| Comment | Team comment management | • Create comment<br>• Delete comment |
| Team | Team management | • Create team<br>• Join team |
| User Management | Admin-only functions | • View user list (admin only)<br>• Delete user (admin only) |



### 5.1.2 Evaluation Metrics

Code coverage is commonly used as a metric for evaluating the quality of test code. Coverage represents the percentage of source code covered by test code and is widely adopted to assess test completeness. In unit testing, coverage is used to measure how much of the source code is tested and whether the tests are executed correctly [7]. However, some studies report no direct correlation between code coverage and software defect rates [19]. Furthermore, in E2E testing, the focus is on software requirements rather than the source code itself. Therefore, even if code coverage is high, it does not necessarily mean that the overall quality of the software is high. For this reason, while Playwright (the framework used in this study) includes a feature to measure code coverage, this study does not use it for evaluation. This study measures the following two evaluation metrics:

- The ratio of successfully compiled test cases
- Functional coverage

These two metrics have been used in previous research on LLM-based unit test improvement and E2E test case generation [7] [9]. ISTQB, an international software testing certification body, defines functional coverage as the ratio of functional requirements addressed by tests relative to all functional requirements of an application. It is widely used as a quality assessment metric for functional testing [20]. Functional coverage is calculated as the ratio of the number of functions covered by tests to the total number of implemented functions within each functional category. This metric is formally defined in Equation (1).

$$Functional\ Coverage = \frac{Number\ of\ functions\ covered\ by\ generated\ tests}{Total\ number\ of\ implemented\ functions} \quad (1)$$

Since E2E testing primarily focuses on functional testing, functional coverage is a suitable metric for evaluating the quality of E2E test code [21]. To verify whether product documentation contributes to improving the quality of generated E2E test code, this study compares three types of documents:

- Product documentation
- Requirement specification documents
- User stories

As discussed in Chapter 3, requirement specification documents and user stories are commonly used as input for LLM-based test generation research and software tools. Additionally, ISTQB's guidelines on functional testing state that functional requirements are often documented in artifacts such as:

- Business requirement specifications
- Epics
- User stories
- Use cases
- Functional specifications

Since these document types can ensure a certain level of test quality, this study compares E2E test code generated from product documentation, requirement specification documents, and user stories [20].

## 5.2 Experimental Results
### 5.2.1 Ratio of Successfully Compiled Tests

Table 4 shows the results of comparing the success rates of compiled test cases. When using product documentation as input, all 53 generated test codes were compiled successfully. All 60 generated test codes were compiled successfully using requirement specification documents. However, when using user stories, only 62 out of 66 generated test codes compiled successfully.

Table 4 List of Application Features

| Method | Total Test Cases | Successful Compilations | Success Rate |
|---|---|---|---|
| Product Documentation | 53 | 53 | 100.00% |
| Requirements Document | 60 | 60 | 100.00% |
| User Stories | 66 | 62 | 93.90% |



### 5.2.2 Functional Coverage

Tables 5 – 10 and Figure 5 present the functional coverage achieved for each document type.

Table 5 Coverage of Authentication Features

| Feature | Product Documentation | Requirements Document | User Stories |
|---|---|---|---|
| User Registration | ✓ | ✓ | ✓ |
| Login | ✓ | ✓ | ✓ |
| Logout | ✓ | – | – |

Table 6 Coverage of Discussion Features

| Feature | Product Documentation | Requirements Document | User Stories |
|---|---|---|---|
| Create Discussion | ✓ | ✓ | ✓ |
| View Discussion | ✓ | ✓ | ✓ |
| Update Discussion | ✓ | – | – |

Table 7 Coverage of Comment Features

| Feature | Product Documentation | Requirements Document | User Stories |
|---|---|---|---|
| Create Comment | ✓ | ✓ | ✓ |
| Delete Comment | ✓ | – | – |

Table 8 Coverage of User Management Features

| Feature | Product Documentation | Requirements Document | User Stories |
|---|---|---|---|
| View User List | ✓ | – | – |
| Delete User (Admin only) | ✓ | – | – |

Table 9 Coverage of Profile Features

| Feature | Product Documentation | Requirements Document | User Stories |
|---|---|---|---|
| View Profile | ✓ | ✓ | – |
| Update Profile | – | ✓ | – |

Table 10 Coverage of Team Features

| Feature | Product Documentation | Requirements Document | User Stories |
|---|---|---|---|
| Create Team | ✓ | ✓ | ✓ |
| Join Team | – | ✓ | – |

Product documentation achieved 100% functional coverage for the authentication, discussion, comment, and user management features. Authentication feature: Covered all functionalities, including user registration, login, and logout, ensuring a complete user authentication flow. Discussion feature: Generated test cases for creating, display, update, and delete operations, fully covering CRUD operations.



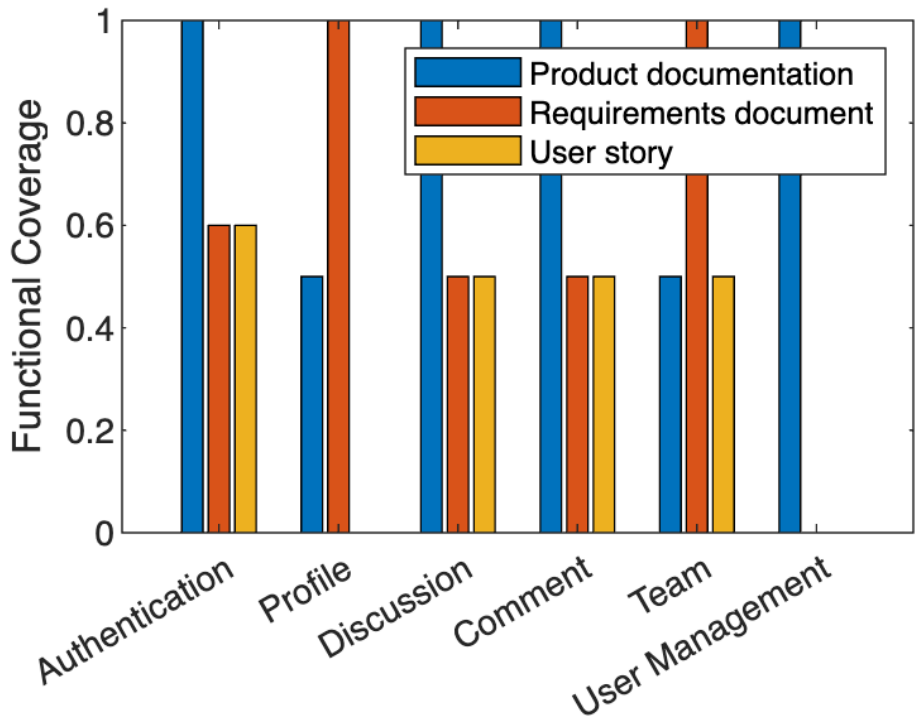

Figure. 5 Comparison of Feature Coverage

Comment feature: Included both creation and deletion operations. User management feature: Covered privileged operations performed by administrators. Requirement specification documents achieved 100% functional coverage for the profile and team management features.

Profile feature: Covered both display and update operations.

Team management feature: Covered team creation and participation operations.

However, only creation and display were covered in the discussion feature, while update and delete operations were not included. Similarly, only creation was covered in the comment feature, while deletion was not included, leading to incomplete coverage in certain functionalities. User stories showed functional coverage of only 50% to 60% across all features, making it the lowest among the three document types.

Authentication feature: Covered user registration and login, but logout was missing.

Discussion feature: Covered create and display, but update and delete operations were missing.

Profile feature: Did not cover either display or update, resulting in a complete lack of coverage.

Observations on Functional Coverage

Product documentation exhibited high functional coverage in features directly related to user operations.

Requirement specification documents achieved reasonable CRUD coverage but lacked coverage for more complex or administrative operations. User stories tended to cover only basic operations, leading to insufficient coverage for detailed functionalities and management operations.

### 5.3 Discussion

The document type does not directly affect the compilation success rate, as compilation only checks for syntax correctness rather than content quality. However, studies suggest that shorter input token lengths reduce LLM inference performance, which might explain why user stories, which contain less detailed content than product documentation and requirement specification documents, resulted in lower-quality generated test code [22].

The high functional coverage achieved in authentication, discussion, comment, and user management features when using product documentation can be attributed to its detailed descriptions of actual operation procedures.



Meanwhile, the requirement specification document achieved high coverage for profile and team management features. This is likely because these features are clearly defined as business logic, and the structured description of functional requirements in the requirement specification document helped derive comprehensive test cases.

Conversely, the low functional coverage of user stories may be due to their focus on user actions and objectives rather than technical details, making it difficult to extract concrete test cases. These results suggest that product documentation plays an effective role in generating E2E tests.

Additionally, combining multiple document types could further enhance test coverage, particularly for complex functionalities. By improving documentation formats or using multiple document types together, it may be possible to generate higher-quality test code.

## 6. Conclusion

This paper proposed and evaluated a method for E2E test code generation using product documentation.

In the experiment, a web application with six key features, including authentication, profile management, and discussion functionalities, was used to compare E2E test code generation based on product documentation, requirement specification documents, and user stories.

The evaluation results demonstrated that when product documentation was used as input, the generated test code achieved high functional coverage, particularly in:

- Authentication feature
- Discussion feature
- Comment feature
- User management feature

Compared to other document types, product documentation enabled the generation of higher-quality test code.

This suggests that the detailed operation procedures and descriptions included in product documentation play a crucial role in generating practical and effective test cases.

However, for team management and profile management features, requirement specification documents led to higher functional coverage than product documentation.

These findings indicate that selecting the appropriate input document type based on feature characteristics can lead to more effective test code generation.

Future research should focus on:

- Combining multiple document types to establish a hybrid approach for test code generation.
- Improving the accuracy of test case generation for more complex functionalities.
- Utilizing image-based information within product documentation, enabling richer data sources for generating test code.

Although automated E2E test code generation remains a challenge, this study suggests that leveraging product documentation as input can potentially improve test code quality. These findings provide valuable insights into efficient software development methods utilizing LLM and documentation. The contributions of this study are expected to serve as a new approach to software test automation and aid future research and development in this field.

## Author contributions statement

T.M. conceived and conducted the experiments, T.M. and H.H. analyzed the results. All authors reviewed the manuscript.

Decisions. In Proceedings of the IEEE/ACM 46th International Conference on Software Engineering, ICSE '24, pp. 1–13, New York, NY, USA, April 2024. Association for Computing Machinery.

[27] Testing Frameworks for Javascript | Write, Run, Debug | Cypress. https://www.cypress.io/. Accessed: January 16, 2025.

[28] Takuya Suemura. Practical Guide to Test Automation: Essential Knowledge and Techniques for Continuously Improving Web Applications, 2024.

[29] Zhiqiang Yuan et al. Evaluating and Improving ChatGPT for Unit Test Generation. Proc. ACM Softw. Eng., Vol. 1, No. FSE, pp. 76:1703–76:1726, July 2024.

[30] Tsuyoshi Yumoto, Toru Matsuodani, and Kazuhiko Tsuda. A Test Analysis Method for Black Box Testing Using AUT and Fault Knowledge. Procedia Computer Science, Vol. 22, pp. 551–560, January 2013.

[31] Takanaka Nakagawa et al. A Mutation-Based Fuzzing Method for Automatic Generation of E2E Tests. Proceedings of the Software Engineering Symposium 2021, Vol. 2021, pp. 303–304, August 2021.